\documentclass[11pt]{article}
\usepackage{moriond,epsfig}
\usepackage{capt-of}

\def\Journal#1#2#3#4{{#1} {\bf #2}, #3 (#4)}

\def\PLB{{\em Phys. Lett.}  B}
\def\PRL{\em Phys. Rev. Lett.}
\def\PRD{{\em Phys. Rev.} D}

\def\bea{\begin{eqnarray}}
\def\eea{\end{eqnarray}}

\newcommand{\bra}{\langle}
\newcommand{\ket}{\rangle}

\newcommand{\subrm}[1]{\mbox{\tiny \rm #1}}

\newcommand{\Br}{\rm Br}
\newcommand{\epe}{\epsilon'/\epsilon\,}

\newcommand{\reeta}{\rm{Re(}\eta_{000}\rm{)}}
\newcommand{\imeta}{\rm{Im(}\eta_{000}\rm{)}}

\newcommand{\pid}{\pi^0_{\subrm{Dalitz}}}
\newcommand{\kl}{K_{\subrm{L}}}
\newcommand{\ks}{K_{\subrm{S}}}

\newcommand{\kspiee}{\ks \to \pi^0 \, e^+ \, e^-}
\newcommand{\klpiee}{\kl \to \pi^0 \, e^+ \, e^-}

\newcommand{\kspipipi}{\ks \to 3 \pi^0}
\newcommand{\klpipipi}{\kl \to 3  \pi^0}

\newcommand{\klpipipic}{\kl \to \pi^+ \pi^- \pi^0}

\newcommand{\kspipi}{\ks \to \pi^+ \pi^-}

\newcommand{\klpizpiz}{\kl \to \pi^0 \pi^0}
\newcommand{\kspizpiz}{\ks \to \pi^0 \pi^0}

\newcommand{\kspipid}{\ks \to \pi^0 \pid}

\newcommand{\kspigg}{\ks \to \pi^0 \gamma \gamma}
\newcommand{\klpigg}{\kl \to \pi^0 \gamma \gamma}

\newcommand{\etazzz}{\eta_{000}}

\newcommand{\bdm}{\begin{displaymath}}
\newcommand{\edm}{\end{displaymath}}
\newcommand{\be}{\begin{equation}}
\newcommand{\ee}{\end{equation}}

\begin{document}
\vspace*{4cm}

\title{NEW RESULTS ON KAON DECAYS FROM NA48}

\author{R.~WANKE}

\address{Institut f\"ur Physik, Johannes Gutenberg-Universit\"at Mainz,\\ D-55099 Mainz, Germany}

\maketitle\abstracts{
New measurements of $\kl$ and $\ks$ decays from the NA48 experiment are presented.
From data taken in the year 2001, the value of the $K_{e3}$ charge asymmetry has been
determined to 
$\delta_L(e) \; = \; (3.317 \pm 0.070_{\subrm{stat}} \pm 0.072_{\subrm{syst}}) \times 10^{-3}$.
From a special high-intensity $\ks$ run period performed in 2000, NA48 has found a limit
of $1.4 \times 10^{-6}$ on the $\kspipipi$ branching ratio and measured
the parameter $\etazzz$.
The same data sample also leads to the first observation of the rare decay $\kspigg$.
All results are still preliminary.
}

\section{Introduction}

\begin{figure}[t]
\begin{center}
\epsfig{file=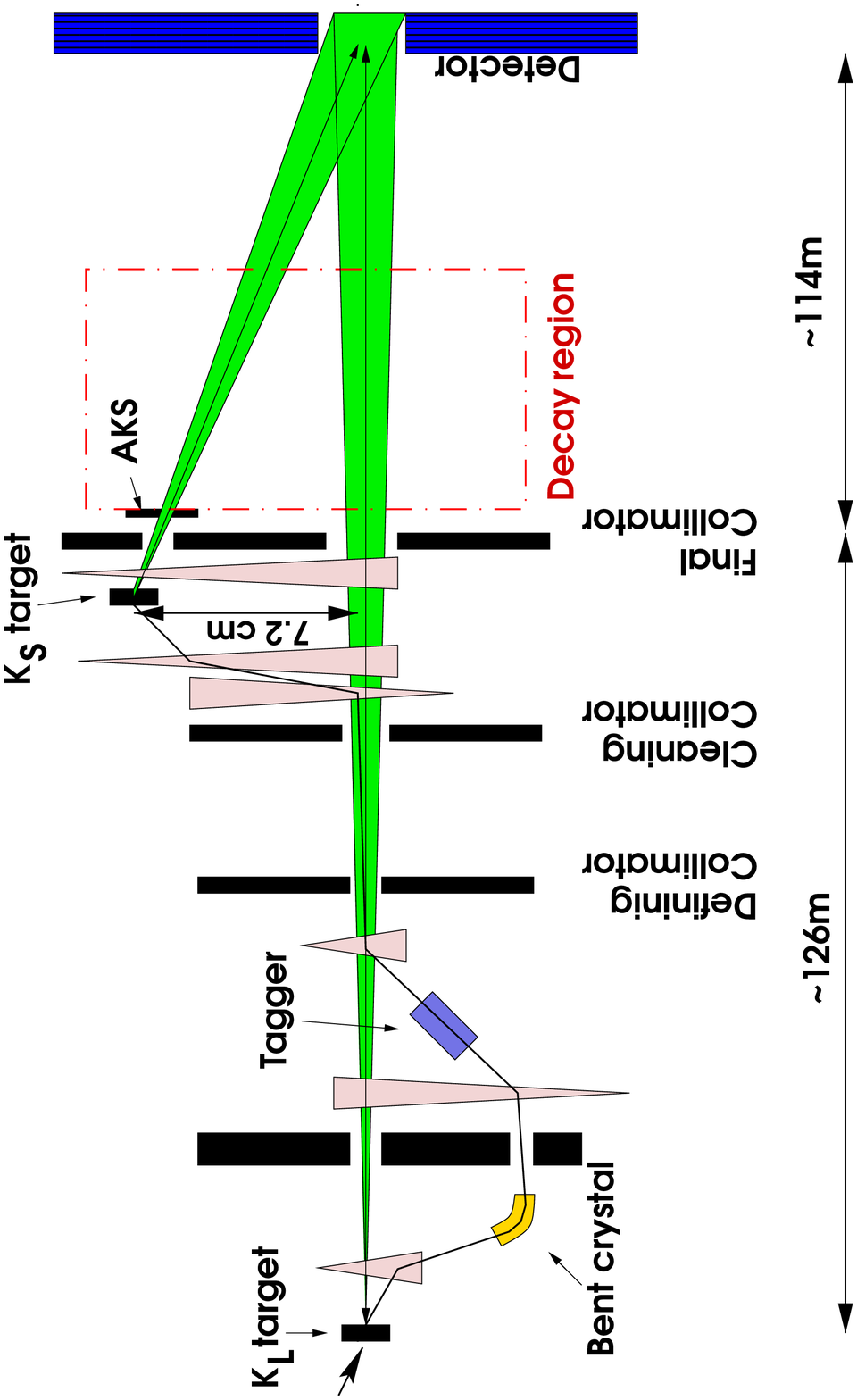,angle=270,width=0.7\linewidth}
\caption{Beam line of the NA48 experiment.}
\label{fig:na48_beams}
\end{center}
\vspace{5mm}
\begin{center}
\epsfig{file=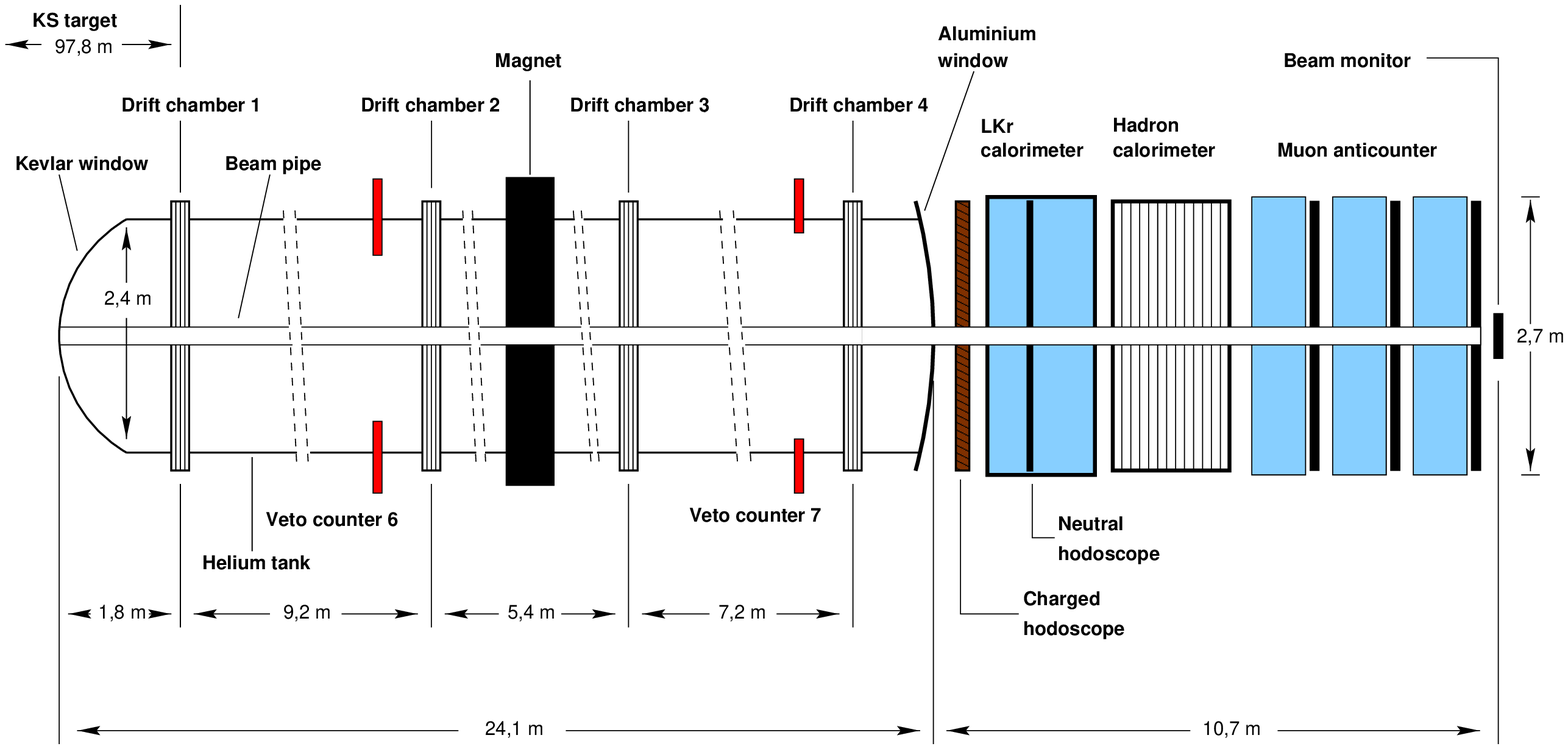,width=0.9\linewidth}
\caption{Side view of the NA48 detector.}
\label{fig:na48_det}
\end{center}
\end{figure}

The NA48 experiment has originally been designed to precisely measure
direct CP violation in the decay of neutral kaons to two pions.
It is a typical fixed target experiment: Neutral kaons and hyperons are produced
by protons interacting with two targets and decay in a roughly 100~m long decay region
(Fig.~\ref{fig:na48_beams}).
One target ({\em far target}) is located 126~m before the begin of the decay region,
a distance large enough that all $\ks$ mesons and neutral hyperons have decayed before
reaching the final collimator.
The other target ({\em near target}) is located only 6~m before the decay region.
Decays observed from this target therefore are predominantly $\ks$ and
and neutral hyperon ($\Lambda$, $\Xi^0$) decays, but may also be $\kl$ decays.
In both neutral beams charged particles are deflected from the beam line by sweeping magnets.
During the CP violation measurement, both targets were used simultaneously,
producing two almost collinear beams of neutral kaons.

Apart from the CP violation measurement, several special run periods have 
been performed for investigations in rare $\kl$, $\ks$, and neutral hyperon decays.
In the year 2000, when due to an accident the drift chambers were not operational,
the run period was split into two halves. The first half with a pure $\kl$ beam from
the far target was used for systematic investigations of the $\epe$ measurement
and for normalization purposes of the subsequently performed high intensity run
using the near target.
Data from this latter period, taken in the second half of the 2000 run period, have been used
for the $\kspipipi$ and $\kspigg$ measurements reported here.

The NA48 detector is shown in Fig.~\ref{fig:na48_det}.
Its main components are a liquid-krypton electro-magnetic calorimeter (LKr)
and a magnetic spectrometer consisting out of two sets of drift chambers
both before and after a dipole magnet and embedded in a helium tank.
The calorimeter has an energy resolution of 
$\Delta E/E  = 3.2\%/\sqrt{E[{\rm GeV}]} \oplus 90 \: {\rm MeV}/E \oplus 0.42\%$
and allows for the precision measurements of neutral decays described below.

For the analyses described here, the magnetic spectrometer 
only is relevant for the $K_{e3}$ charge asymmetry measurement.
In the 2000 run period it was replaced by a vacuum tank
to minimize multiple scattering.

\section{Measurement of the $K_{e3}$ Charge Asymmetry}

The charge asymmetry in semileptonic $\kl$ decays is an important observable
for CP violation.
For $\kl \to \pi e \nu$ ($K_{e3}$) decays it is defined as
\[
\delta_L(e) \; = \; \frac{\Gamma(\kl \to \pi^- e^+ \nu)-\Gamma(\kl \to \pi^+ e^- \bar{\nu})}
                         {\Gamma(\kl \to \pi^- e^+ \nu)+\Gamma(\kl \to \pi^+ e^- \bar{\nu})}
\]
and equals to $2 \times {\rm Re}(\epsilon)$ in case of CPT symmetry.
Very accurate measurements of the charge asymmetry have already be performed in the 1970's.
However, the most precise value comes from a recent measurement of KTeV which yields 
$\delta_L(e) = (3.322 \pm 0.058 \pm 0.047) \times 10^{-3}$~\cite{bib:ke3_ca_ktev}.

\begin{figure}
\begin{minipage}{0.6\linewidth}
\begin{center}
\epsfig{file=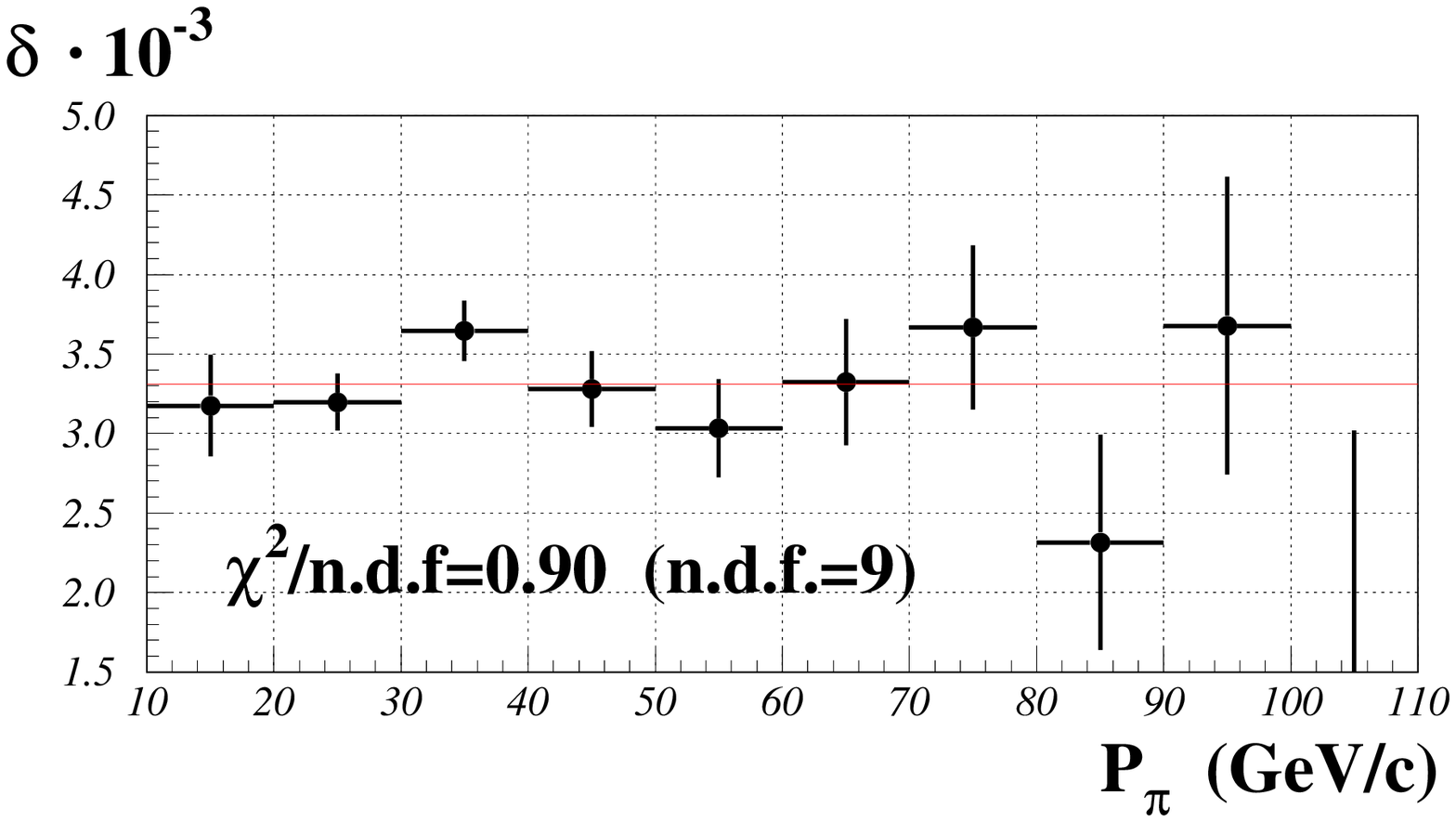,width=0.89\linewidth}
\vspace{-3mm}
\caption{Fit results for $\delta_L(e)$ in bins of the pion momentum.}
\vspace{5mm}
\label{fig:ke3}
\end{center}
\end{minipage}
\hfill
\begin{minipage}{0.38\linewidth}
\begin{center}
\vspace{3mm}
\begin{tabular}{l|r}
               & \multicolumn{1}{c}{Correction}  \\ 
               & \multicolumn{1}{c}{in $10^{-5}$}  \\ \hline
Trigger        & $+26.2 \, \pm \, 6.0$ \\
Punch through  & $ -1.4 \, \pm \, 3.5$ \\
Pion ID        & $-17.1 \, \pm \, 2.4$ \\
Acceptance     & $         \pm \; 0.5$ \\
Background     & $         \pm \; 0.5$ \\ \hline \hline
Total          & $ +7.7 \, \pm \, 7.2$ \\
\end{tabular}
\vspace{2mm}
\captionof{table}{Corrections and systematic uncertainties on $\delta_L(e)$.}
\label{tab:ke3_syst}
\end{center}
\end{minipage}
\end{figure}

In the run period of 2001, alongside the $\pi \pi$ events used for the final $\epe$ measurement,
the NA48 experiment has collected about $2 \times 10^8$ $K_{e3}$ events in a special
trigger stream.
During the data taking, the 
sign of the magnet current was swapped several times to reduce systematics.
All backgrounds have been estimated to be negligible.
The main systematic uncertainties arise from asymmetries in the
energy deposition of the differently charged pions in the LKr calorimeter. 
This results in asymmetries in the efficiency of the total-energy trigger
and in the pion identification when comparing the LKr-measured energy with the
momentum measured in the spectrometer.
Both these effects are strongly momentum dependent. 
They are corrected for by the use of data from control triggers and
from well-known control samples as $\klpipipic$ or $\kspipi$.
After the correction, no momentum dependency of the result can be seen (Fig.~\ref{fig:ke3}).

A summary of the applied corrections and the systematic uncertainties
is given in Tab.~\ref{tab:ke3_syst}. After the corrections, the (preliminary)
result on the charge asymmetry is
\[
\delta_L(e) \; = \; (3.317 \pm 0.070_{\subrm{stat}} \pm 0.072_{\subrm{syst}}) \times 10^{-3},
\]
in very good agreement with the KTeV result quoted above.

\section{Search for the CP violating Decay $\kspipipi$ and Measurement of $\etazzz$}

The decay $\kspipipi$ is purely CP violating in complete analogy, but with reversed CP values, 
to $\klpizpiz$.
The amplitude ratio compared to the corresponding $\kl$ decay is expected to be 
\[
\etazzz \; \equiv \; \frac{A(\kspipipi)}{A(\klpipipi)}
        \; = \; \epsilon + i \, \frac{{\rm Im}(A_1)}{{\rm Re}(A_1)}.
\]
While the real part is fixed by CPT conservation
the imaginary part depends on the isospin 1 amplitude $A_1$ and may differ from 
Im$(\epsilon)$.
Experimentally, $\kspipipi$ has never been observed. The parameter $\etazzz$
has been measured to $\reeta = 0.18 \pm 0.15$ and $\imeta = 0.15 \pm 0.20$
by CPLEAR~\cite{bib:ks3pi0_cplear}. In addition, the SND experiment
has set a limit on the branching fraction of  $1.4 \times 10^{-5}$
at 90\% confidence level~\cite{bib:ks3pi0_snd}.

In NA48 $\kl$ and $\ks$ mesons are produced in equal amounts at the target. The dependence
of the $K^0/\overline{K^0} \to 3 \pi^0$ intensity as function of proper time therefore is given by
\[
I_{3\pi^0}(t) \propto \! \! \underbrace{\raisebox{0mm}[0mm][2mm]{$e^{- \Gamma_L t}$}}_{\textstyle \raisebox{-1mm}{$\kl$ decay}} \! \! 
                    + \: \underbrace{\raisebox{0mm}[0mm][2mm]{$|\etazzz|^2 \, e^{- \Gamma_S  t}$}}_{\textstyle \raisebox{-1mm}{$\ks$ decay}}
                    \: + \: \underbrace{\raisebox{0mm}[0mm][2mm]{$2 \, D(p) \left( {\rm Re}(\etazzz) \cos{\Delta m  t}  -
                                                        {\rm Im}(\etazzz) \sin{\Delta m t} \right)
                                             \, e^{ - \frac{1}{2}(\Gamma_S + \Gamma_L) \, t}$}}_{\textstyle \raisebox{-1mm}{$\kl$-$\ks$ interference}}
\]
with the momentum dependent dilution
$D(p) = (N(K^0) - N(\overline{K^0})/(N(K^0) - N(\overline{K^0})$.
In NA48, this production asymmetry is on average about 0.35 with an almost linear
dependency from the kaon momentum.
While it is hopeless to observe the pure $\ks$ contribution above the $\kl$ decay,
the $\kl \ks$ interference is suppressed by only the first order in $\etazzz$.

\begin{figure}[t]
\begin{center}
\epsfig{file=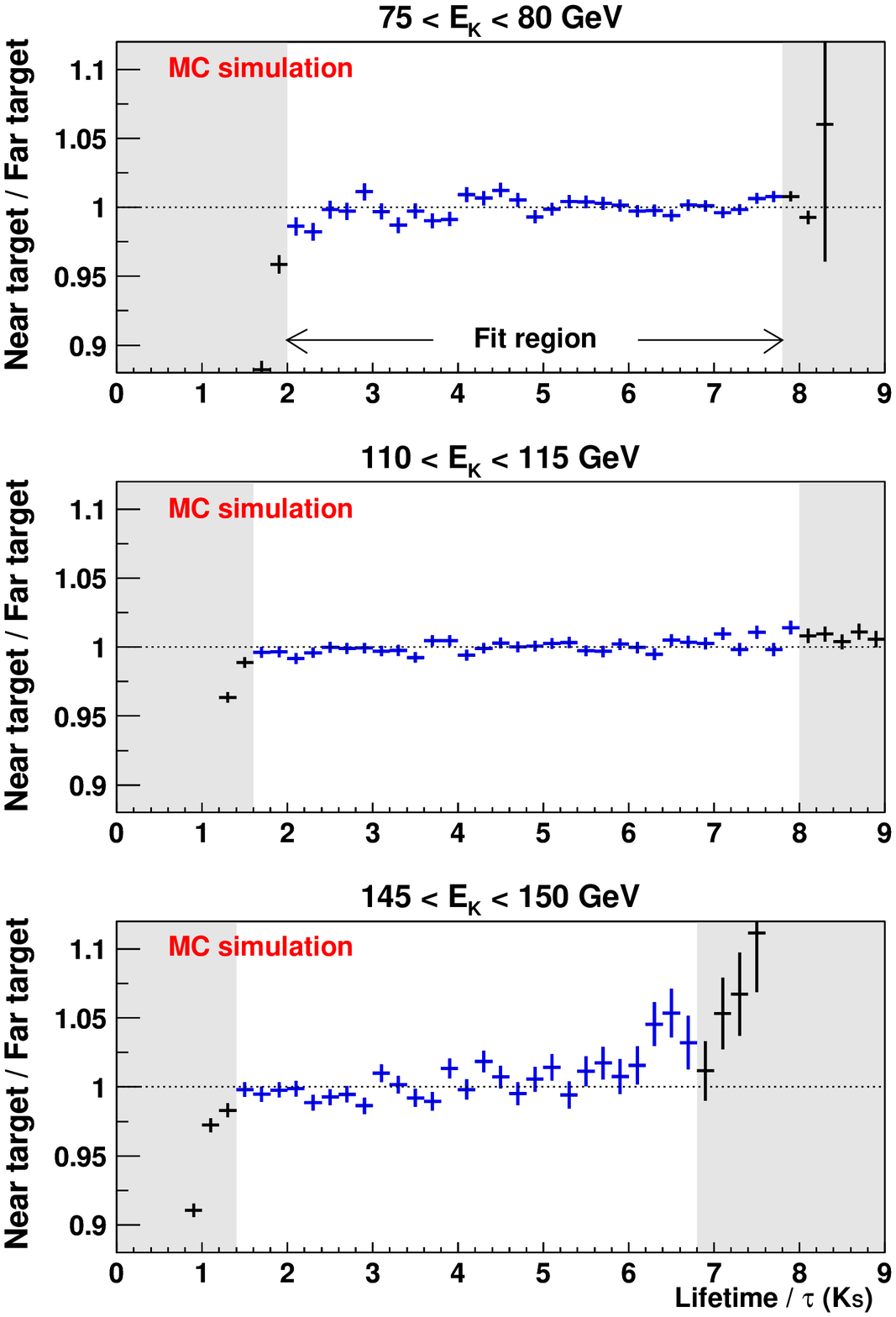,width=0.45\linewidth}
\hfill
\epsfig{file=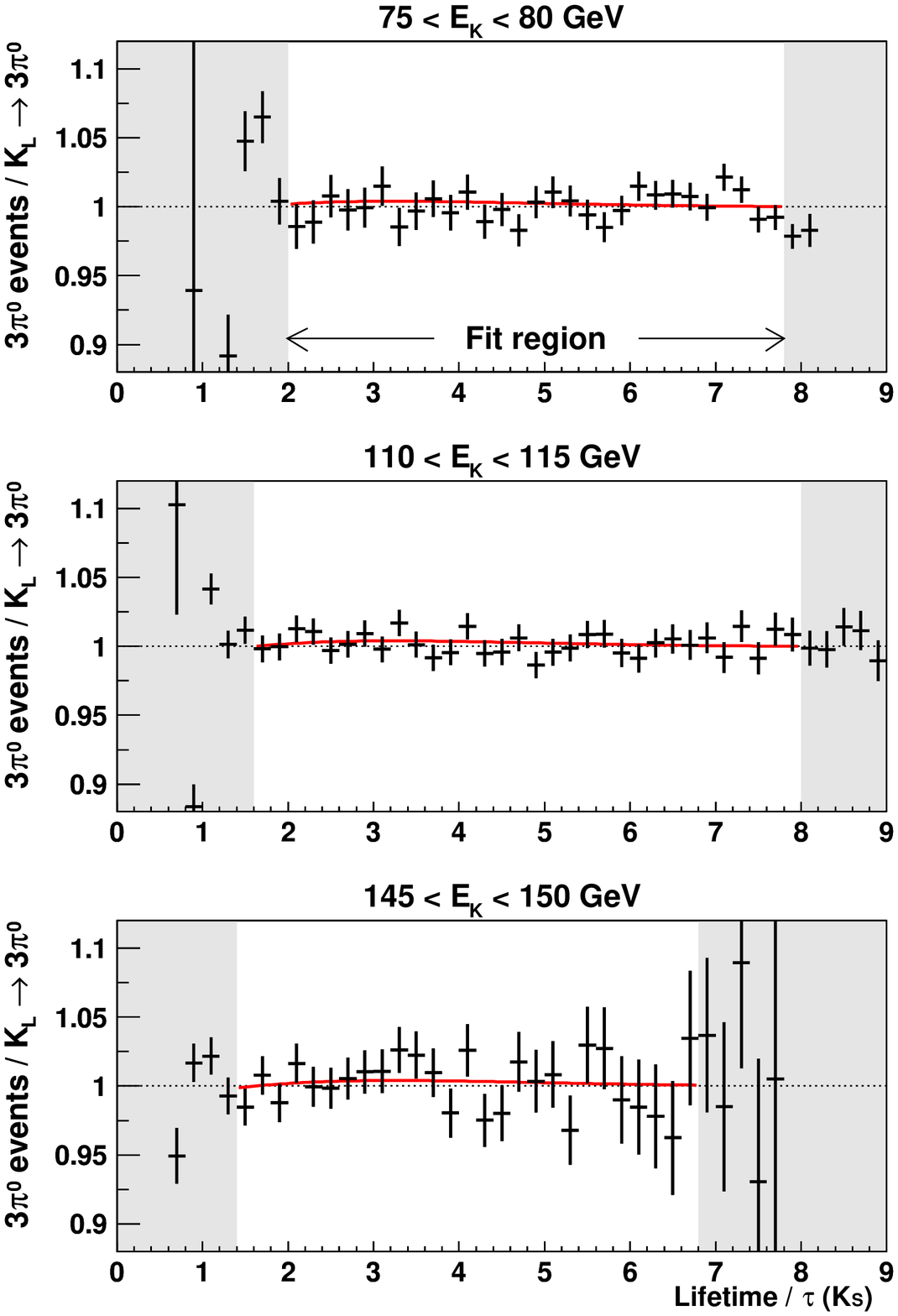,width=0.45\linewidth}
\caption{Ratios of $3 \pi^0$ events for different energy bins. Left: 
$\klpipipi$ Monte Carlo simulation from near and far target.
Right: Near-target data normalized to geometry-corrected far-target
$\klpipipi$ data.}
\label{fig:ks3pi0_fit}
\end{center}
\end{figure}

\begin{figure}[th]
\vspace{-4mm}
\begin{minipage}{0.5\linewidth}
\begin{center}
\vspace{23mm}
\renewcommand{\arraystretch}{1.1}
\begin{tabular}{l|c|c}
                              & $\reeta$       & $\imeta$       \\ \hline
Acceptance                    & $\pm \, 0.003$ & $\pm \, 0.008$ \\
Accidental activity           & $\pm \, 0.001$ & $\pm \, 0.006$ \\
Energy scale                  & $\pm \, 0.001$ & $\pm \, 0.001$ \\
$K^0 \overline{K^0}$ dilution & $\pm \, 0.003$ & $\pm \, 0.004$ \\
Fit                           & $\pm \, 0.001$ & $\pm \, 0.002$ \\ \hline \hline
Total:                        & $\pm \, 0.005$ & $\pm \, 0.011$ \\
\end{tabular}
\renewcommand{\arraystretch}{1.0}
\vspace{4mm}
\captionof{table}{Systematic uncertainties on $\etazzz$.}
\label{tab:ks3pi0_syst}
\end{center}
\end{minipage}
\hfill
\begin{minipage}{0.45\linewidth}
\begin{center}
\epsfig{file=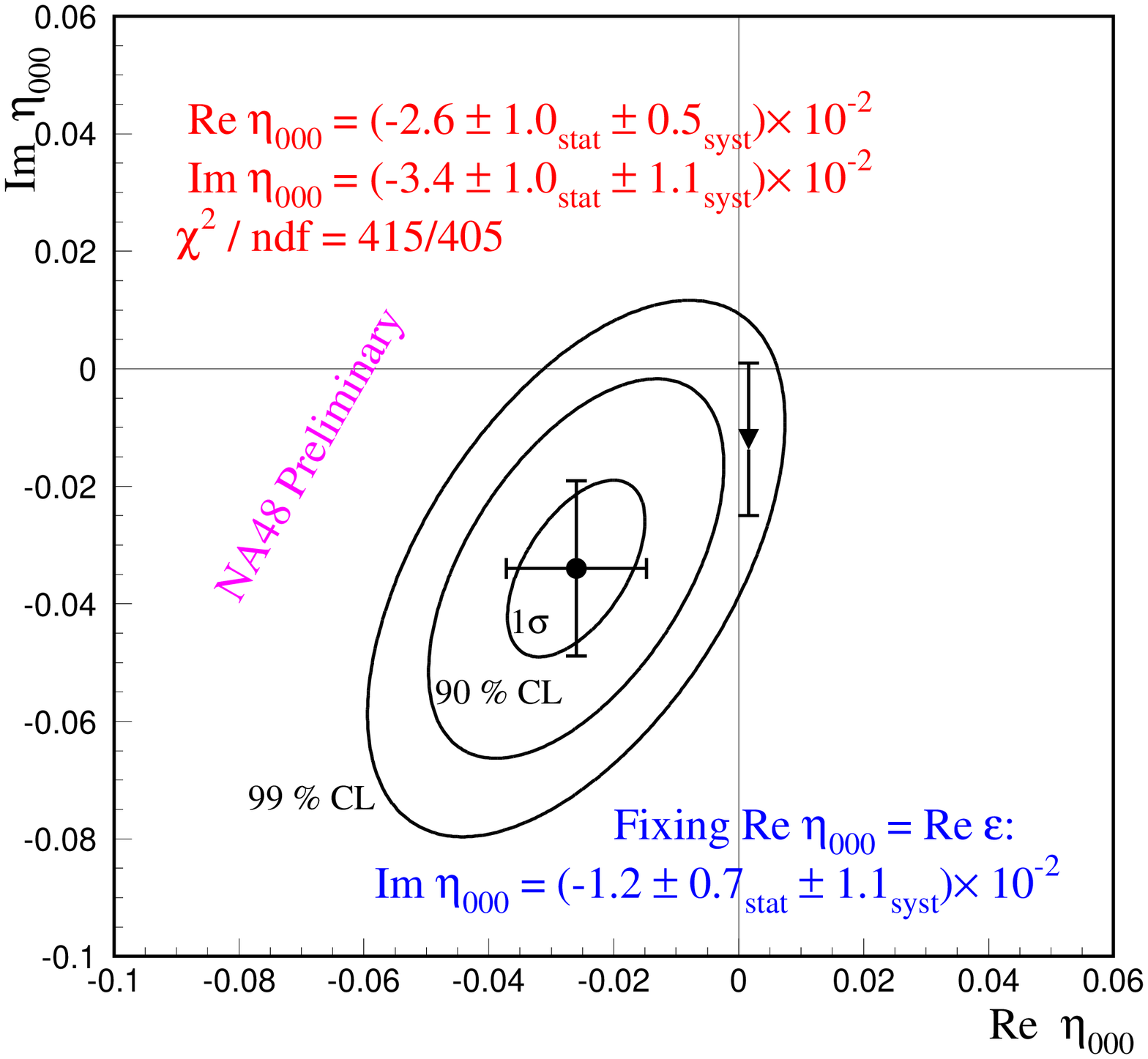,width=0.96\linewidth}
\vspace{-4mm}
\caption{Fit result for $\etazzz$ (preliminary).}
\vspace{-2mm}
\label{fig:ks3pi0_result}
\end{center}
\end{minipage}
\vspace{-1mm}
\end{figure}

For the $\etazzz$ measurement data from the 2000 near-target
run period were used. From these data about $6.5 \times 10^6$ $3\pi^0$ events were
selected with practically negligible background. 
To be as independent from Monte Carlo simulation as possible data from the far-target
run period of the same year were used for normalization to $\klpipipi$.
In this way, due to the almost identical geometry of near- and far-target beams, only residual effects had to be corrected for by Monte Carlo simulation
(see Fig.~\ref{fig:ks3pi0_fit} left).
To also be independent of the correct modeling of the different kaon energy spectra in the simulation,
the fit to the proper time distribution of the $3 \pi^0$ events was performed in bins of energy,
leaving all normalizations free in the fit.
The fit result is $\reeta = -0.026 \pm 0.010$ and $\imeta = -0.034 \pm 0.010$
with a correlation coefficient of 0.8.
The systematic uncertainties are dominated by uncertainties in the detector acceptance,
the accidental activity, and the $K^0 \overline{K^0}$ dilution (see Tab.~\ref{tab:ks3pi0_syst}).
The complete result, which is still preliminary, then is
\[
\reeta = -0.026 \pm 0.010_{\subrm{stat}} \pm 0.005_{\subrm{syst}}
\quad \mathrm{and} \quad
\imeta = -0.034 \pm 0.010_{\subrm{stat}} \pm 0.011_{\subrm{syst}}.
\]

The corresponding confidence limits are shown in Fig.~\ref{fig:ks3pi0_result}.
The result is consistent with 0 with roughly 5\% probability.
Turning this result into an upper limit on the branching fraction, we get
\[
\Br(\kspipipi) < 1.4 \times 10^{-6} \quad {\rm at \: 90\% \: CL,}
\]
which is one order of magnitude below the previous best limit.

Assuming CPT conservation, which fixes $\reeta$ to
${\rm{Re(}\epsilon\rm{)}} = 1.6 \times 10^{-3}$, we receive
\[
\imeta|_{{\rm Re}(\etazzz) = {\rm Re}(\epsilon)} = -0.012 \pm 0.007_{\subrm{stat}} \pm 0.011_{\subrm{syst}}
\]
and $\Br(\kspipipi)|_{{\rm Re}(\etazzz) = {\rm Re}(\epsilon)} < 3.0 \times 10^{-7}$ at 90\% CL.

Finally, this result improves the limit on CPT violation via the
Bell-Steinberger relation. Using unitarity, this relation connects the CPT violating phase
$\delta$ with the CP violating amplitudes of the various $\kl$ and $\ks$ decays~\cite{bib:bellsteinberger}.
So far, the limiting quantity has been the precision of $\etazzz$. This new result, added to the measurements
of the other $\eta$ parameters~\cite{bib:pdg,bib:k3pi_cplear,bib:cpt_cplear},
improves the accuracy on Im$(\delta)$ by about 40\% to Im$(\delta) = (-1.2 \pm 3.0) \times 10^{-5}$,
now limited by the knowledge of $\eta_{+-}$.
Assuming CPT conservation in the decay, this can be converted to a new limit on the
$K^0\overline{K^0}$ mass difference 
of $m_{K^0} - m_{\overline{K^0}} = (-1.7 \pm 4.2) \times 10^{-19}$~GeV/$c^2$.

\section{First Observation of the Decay $\kspigg$}

\begin{figure}[t]
\begin{minipage}{0.31\linewidth}
\begin{center}
\epsfig{file=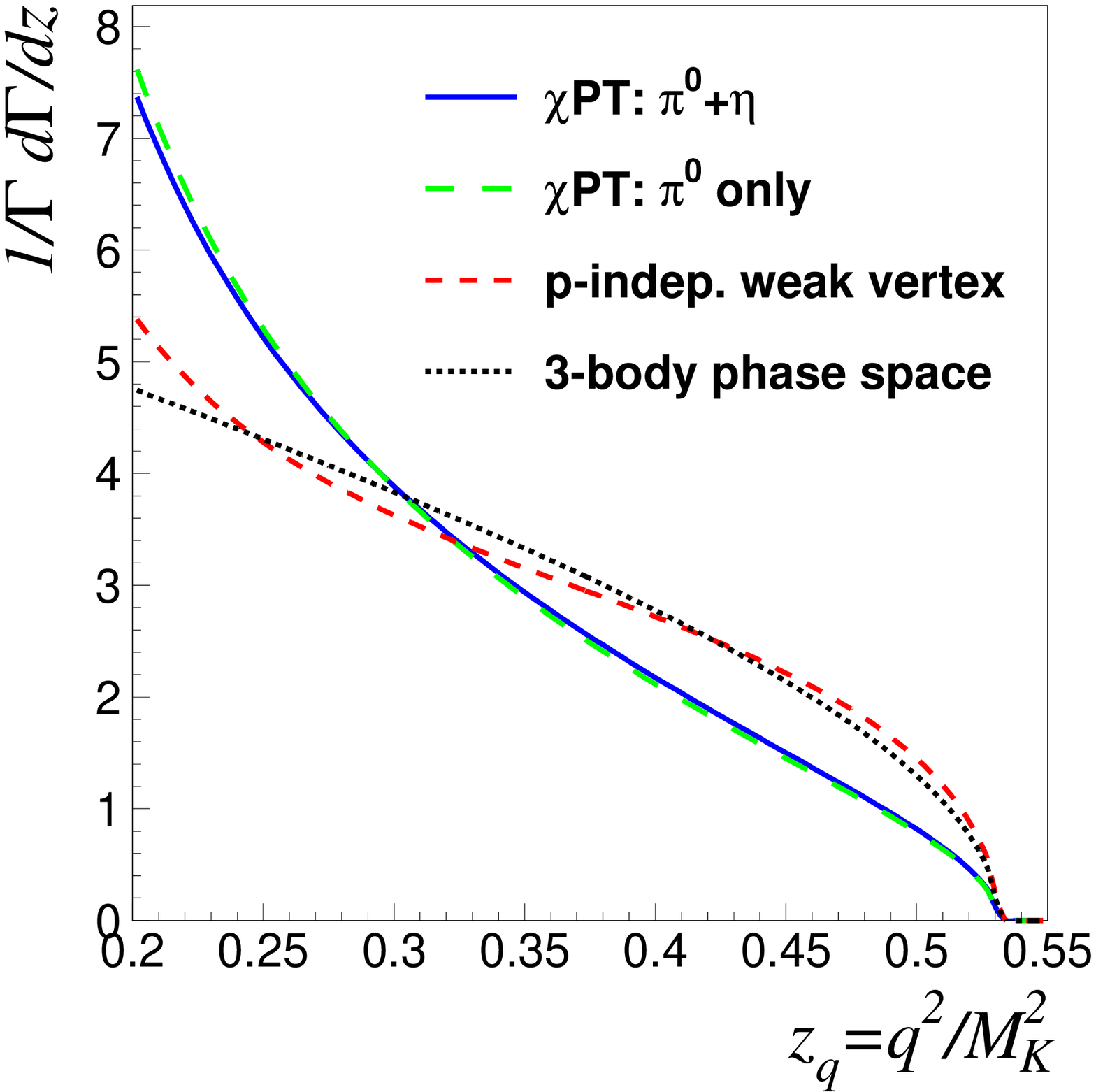,width=\linewidth}
\vspace{-7mm}
\caption{Theoretical predictions for $z = q^2/m_K^2$ in $\kspigg$ from 3-body phase space and ChPT.}
\label{fig:kspigg_zpred}
\end{center}
\end{minipage}
\hspace*{0.02\linewidth}
\begin{minipage}{0.31\linewidth}
\begin{center}
\epsfig{file=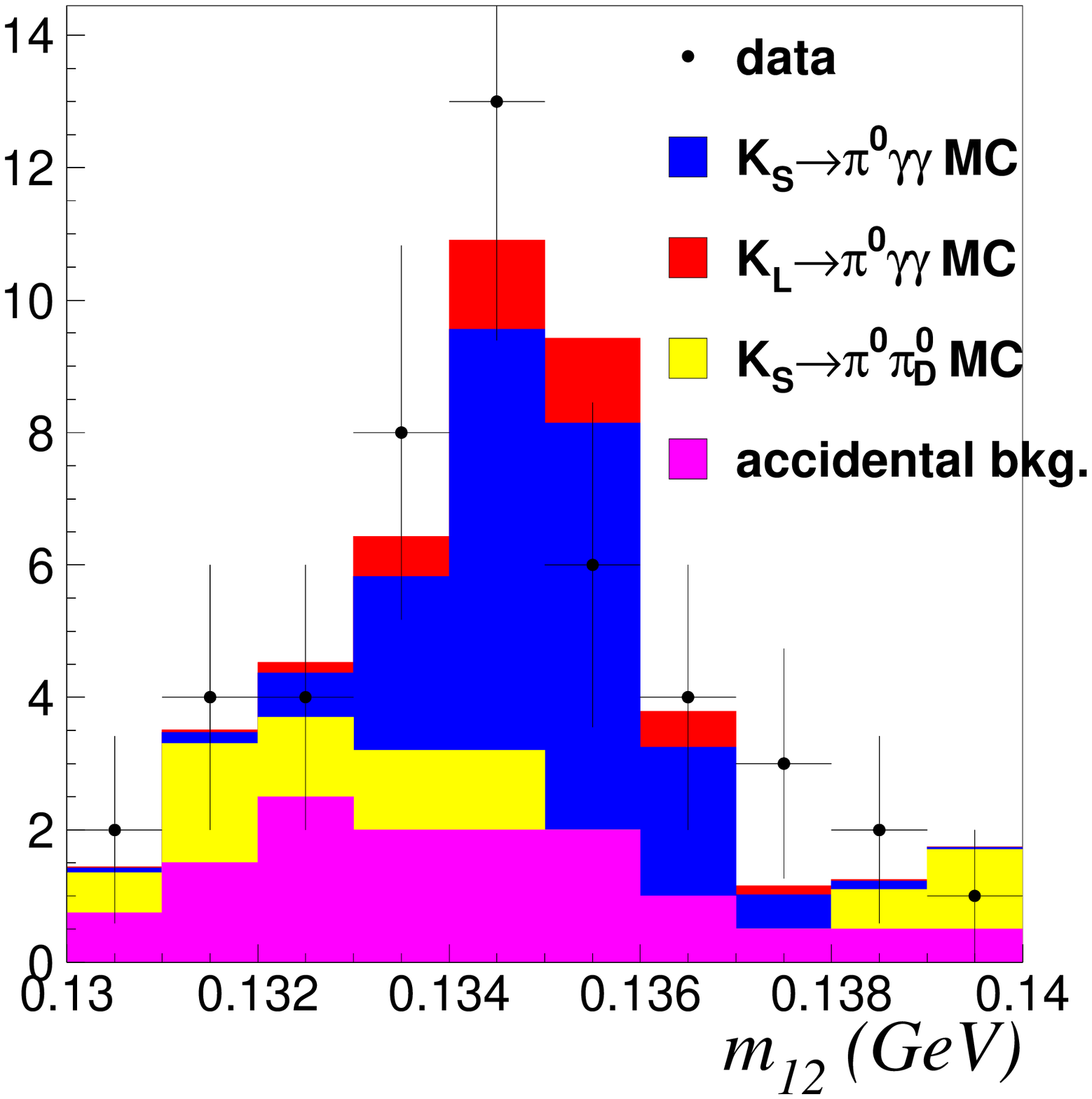,width=\linewidth}
\vspace{-7mm}
\caption{Invariant $m_{\gamma_1 \gamma_2}$ mass for the selected $\kspigg$ candidates
with estimated background.}
\label{fig:kspigg_sig}
\end{center}
\end{minipage}
\hspace*{0.02\linewidth}
\begin{minipage}{0.31\linewidth}
\begin{center}
\epsfig{file=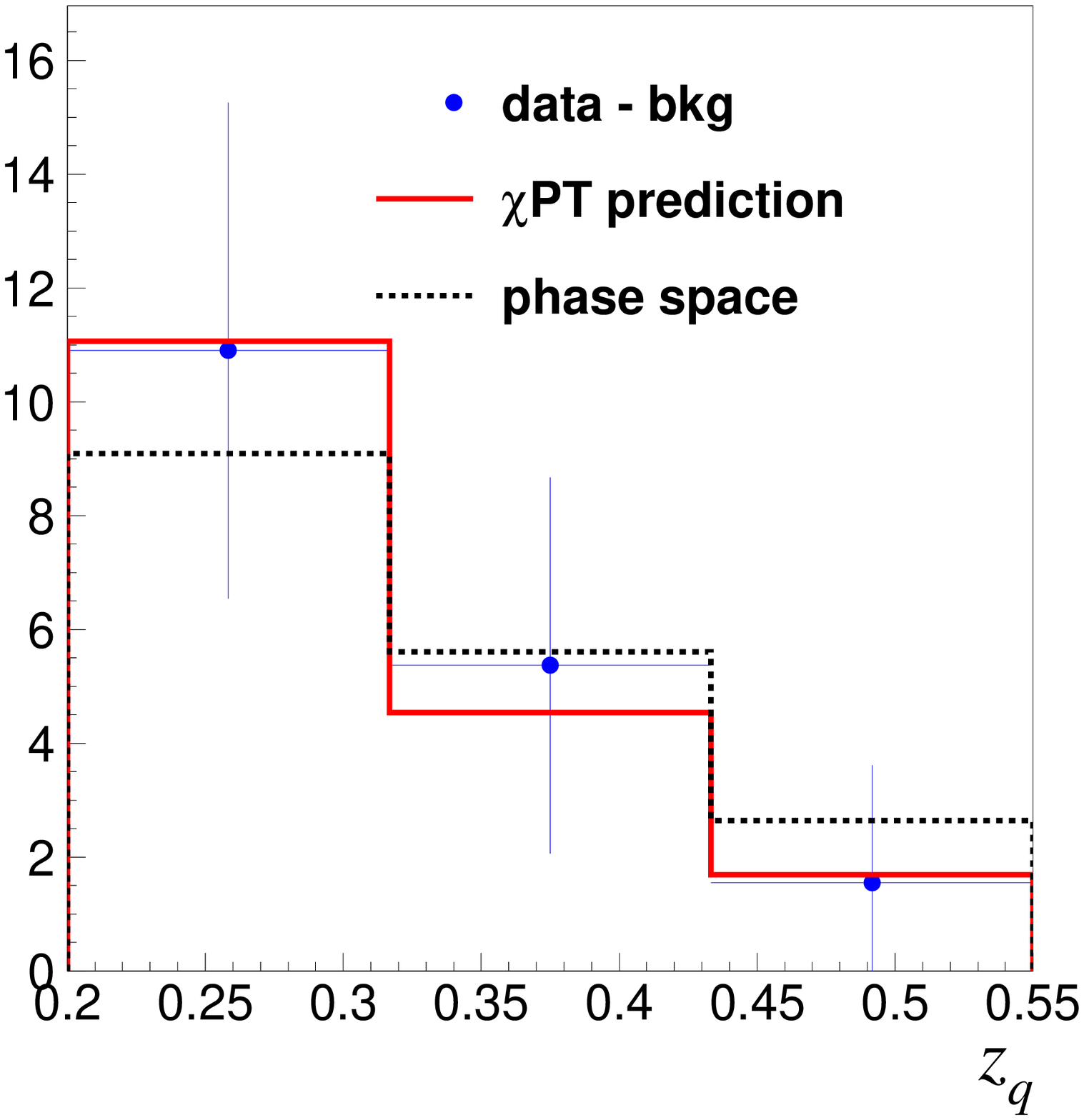,width=\linewidth}
\vspace{-7mm}
\caption{Measured $z = q^2/m_K^2$ for the fitted $\kspigg$ signal and predictions.}
\label{fig:kspigg_zdata}
\end{center}
\end{minipage}
\end{figure}

The decay $\kspigg$ has been investigated by using Chiral Perturbation Theory (ChPT)~\cite{bib:kspigg_ecker}.
A comparison of the predicted spectra with pure phase space is shown in Fig.~\ref{fig:kspigg_zpred}
for $z \equiv q^2/m_K^2 = m_{\gamma_3 \gamma_4}^2/m_K^2 > 0.2$ far away from the pion pole.
The branching fraction is predicted to
Br$(\kspigg)|_{z>0.2} = 3.8 \times 10^{-8}$.
Experimentally, the decay $\kspigg$ had not yet been observed.
The best limit has been set only recently by NA48, using data from a test run
in 1999~\cite{bib:kspigg99}.

The new NA48 analysis uses
the data from the high-intensity near-target run period of the year 2000.
After applying all selection criteria 31~signal candidates remain.
Although great care has been taken to suppress all sources of background,
the selected sample still contains estimated $13.6 \pm 2.7$ background events.
They consist mostly of accidental beam activity ($7.4 \pm 2.4$ events),
which was estimated using time sidebands. Other large sources are irreducible
but well-known $\klpigg$ events and misidentified $\kspipid$ decays.
The distribution in Fig.~\ref{fig:kspigg_sig} shows the reconstructed invariant $\pi^0$ mass of
the selected events together with fitted signal and background expectations.

For computing the branching fraction the signal sample is normalized to $\kspizpiz$ decays.
The result is
\[
{\rm Br}(\kspigg)|_{z>0.2} = (4.9 \pm 1.6_{\subrm{stat}} \pm 0.8_{\subrm{syst}} ) \times 10^{-8}.
\]
The systematic error is dominated by uncertainties in the background subtraction.
The probability of a background fluctuation has been estimated to be less than $9 \times 10^{-4}$.
The result is compatible with the prediction quoted above.
However, the statistics are still too low to conclude from the
$m_{\gamma_3 \gamma_4}$ spectrum to the structure of the weak vertex (Fig.~\ref{fig:kspigg_zdata}).

\section{A Look into the Future}

After having finished the $\epe$ program, the NA48 collaboration performs two further programs:
For the NA48/1 experiment~\cite{bib:add2}
the whole 2002 run period was devoted for investigations of rare $\ks$ and neutral hyperon
decays with a high-intensity $\ks$ beam. Among many other interests, the prime goals of this 
run are the measurement of the rare decay $\kspiee$, which determines the indirectly CP
violating amplitude of $\klpiee$, and the precision measurement of
$\Xi^0$ beta decays.
A first look into the 2002 data shows more than 9000 $\Xi^0 \to \Sigma^+ e^- \bar{\nu_e}$ candidates
over very little background. Also a first signal of the so far unobserved channel
$\Xi^0 \to \Sigma^+ \mu^- \bar{\nu_\mu}$ is seen.

In the year 2003 and possibly 2004 the experiment NA48/2~\cite{bib:add3} will investigate
$K^\pm$ decays. For this experiment the beam line is greatly modified and a beam spectrometer
for $K^\pm$ momentum determination is added.
In total, about $3 \times 10^{11}$ $K^\pm$ decays are expected in 2003 in the fiducial volume.
The main goal of this experiment is the search for CP violation 
in the slope of $K^\pm \to \pi^+ \pi^- \pi^\pm$ Dalitz plot, which might be as as large
as $10^{-4}$. 
Other goals of the NA48/2 experiment are the measurement of the quark condensate
$\bra 0 | q \bar{q} | 0 \ket$ in $K_{e4}$ decays ($K^+ \to \pi^+ \pi^- e^+ \nu_e$),
an absolute measurement of Br$(K^+ \to \pi^0 e^+ \nu_e)$ for determination of $V_{us}$,
and many investigations of rare $K^+$ decays such as $K^+ \to \pi^+ \gamma \gamma$
or $K^+ \to l^+ l^- l^+ \nu_l$.

\end{document}